\documentstyle[preprint,aps,12pt,tighten]{revtex}

\begin{document}
\title{Decomposition Theory of Spin Connection and Topological Structure of
Gauss-Bonnet-Chern Theorem on Manifold With Boundary}
\author{Sheng Li\thanks{%
Email: lisheng@itp.ac.cn}}
\address{CCAST World Lab, Academia Sinica, Beijing 100080, P. R. China}
\address{Institute of Theoretical Physics, Academia Sinica, Beijing 100080, P. R.
China}
\author{Yishi Duan\thanks{%
Email: ysduan@lzu.edu.cn}}
\address{Institute of Theoretical Physics, Lanzhou University, Lanzhou 730000, P. R.
China}
\date{20 January, 1999}
\draft
\maketitle

\begin{abstract}
The index theorem of Euler-Poincar\'{e} characteristic of manifold
with boundary is given by making use of the general decomposition
theory of spin connection. We shows the sum of the total index of
a vector field $\phi $ and half the total of the projective vector
field of $\phi $ on the boundary equals the Euler-Poincar\'{e}
characteristic of the manifold. Detailed discussion on the
topological structure of the Gauss-Bonnet-Chern theorem on
manifold with boundary is given. The Hopf indices and Brouwer
degrees label the local structure of the Euler density.
\end{abstract}


\section{Introduction}

Gauss-Bonnet-Chern($GBC$) theorem is one of the most significant results in
differential geometry, which relates the curvature of the compact and
oriented even-dimensional Riemannian manifold ${\bf M}$ with an important
topological invariant, the Euler-Poincar\'{e} characteristic $\chi {({\bf M})%
}$. An elegantly intrinsic proof of the theorem was given by professor Chern%
\cite{Cher1}, whose instructive idea is to work on the sphere bundle $S({\bf %
M})$. An useful summary and some historical comments on the $GBC$ theorem
are given respectively by Kabayashi, Nomizu\cite{Koba}, and Spivak\cite{Spiv}%
. A detailed review of Chern's proof of the $GBC$ theorem is presented in
Ref.\cite{Dowk}.

A great advance in this field was the discovery of the relationship between
supersymmetry and the index theorem, which includes the derivation of the $%
GBC$ theorem via supersymmetry and the path integral techniques as presented
by Alvarez-Gaum\'{e} et al\cite{Alva}. In topological quantum field theory
which was initiated by Witten\cite{Witt}, the $GBC$ theorem can be derived
by means of Morse theory\cite{Laba}. On the physics side, the optical Berry
phase is a direct result of the Gauss-Bonnet theorem\cite{Ryde} and the
black hole entropy emerges as the Euler class through dimensional
continuation of the Gauss-Bonnet theorem\cite{Bana}.

In this paper a generalized Hopf theorem on manifolds with boundaries is
given. It says the Euler-Poincar\'{e} characteristic $\chi $ of a manifold $M
$ with boundary $\partial M$ equals the sum of the total index the vector
field $\phi $ on $M$ and half the total index of the projection vector of $%
\phi $ on the boundary $\partial M$. A detailed discussion on the local
structure of the $GBC$ density (the Euler-Poincar\'{e} characteristic $\chi (%
{\bf M})$ density) shows that only the zeroes of $\phi $ and the zeroes of
the projective vector of $\phi $ on $\partial M$ contribute to $\chi ({\bf M}%
)$. We also show that the Brouwer degrees and Hopf indices label the local
structure of the $GBC$ density. The direct relationship between the
Euler-Poincar\'{e} characteristic and winding number is given.

The generalized decomposition theory of spin connection using in
this paper is an useful tool in the research of topology in gauge
theory. This theory has been effectively used in the magnetic
monopole problem\cite{Hoof,Duan1}, the topological gauge theory of
dislocation and disclination in condensed matter
physics\cite{Duan5}, the space-time defects\cite{Duan} and
Gauss-Bonnet-Chern($GBC$) theorem\cite{Duan6,lisheng}. The
essential feature of the decomposition theory shows that the
connection can be decomposed and possess inner structures. In this
paper, we give a general decomposition and inner structure of the
spin connection of the group $SO(N)$ using $N$ orthonormal vectors
via Clifford algebra. The curvature under this decomposition is
found to be a generalized function. The singular points of the
curvature are the sources of the non-triviality of the manifold.

This paper is arranged as follows. In section 2, we deduce the general
decomposition and inner structure of spin connection via $N$ orthonormal
vectors. Using the formula given in section2, the $GBC$ form is expressed in
terms of an unit vector in section 3, In section 4 and 5, the topological
structure of the Chern density on manifold without boundary and without
boundary are given respectively. In section 6, we present a short summary.

\section{The decomposition theory of spin connection}

Let ${\bf M}$ be a compact and oriented $N$-dimensional Riemannian manifold
and $P(\pi ,{\bf M},G)$ be a principal bundle with the structure group $%
G=SO(N)$. A smooth vector field $\phi ^a$ $(a=1,2,\cdots ,N)$ can be found
on the base manifold ${\bf M}$ (a section of a vector bundle over ${\bf M}$%
). We define a unit vector $n$ on ${\bf M}$ as
\begin{equation}
n^a=\phi ^a/||\phi ||\qquad {}a=1,2,\cdots ,N  \label{f201}
\end{equation}
\[
||\phi ||=\sqrt{\phi ^a\phi ^a},
\]
in which the superscript $``a"$ is the local orthonormal frame index.

In fact $n$ is identified as a section of the sphere bundle over ${\bf M}$
(or a partial section of the vector bundle over ${\bf M}$). We see that the
zeroes of $\phi $ are just the singular points of $n$. Since the global
property of a manifold has close relation with zeroes of a smooth vector
fields on it, this expression of the unit vector $\vec{n}$ is a very
powerful tool in the discussion of the global topology. It naturally
guarantees the constraint
\begin{equation}
n^an^a=1,  \label{f202}
\end{equation}
and

\begin{equation}
n^adn^a=0.  \label{f203}
\end{equation}
The covariant derivative $1$-form of $n^a$ is
\begin{equation}
Dn^a=dn^a-\omega ^{ab}n^b,  \label{f204}
\end{equation}
and the curvature $2$-form is
\begin{equation}
F^{ab}=d\omega ^{ab}-\omega ^{ac}\wedge \omega ^{cb},  \label{f205}
\end{equation}
where $\omega ^{ab}$ is the spin connection $1$-form:
\begin{equation}
\omega ^{ab}=\omega _\mu ^{ab}dx^\mu \qquad \omega ^{ab}=-\omega ^{ba},
\label{f206}
\end{equation}
and
\begin{equation}
F^{ab}=\frac 12F_{\mu \nu }^{ab}dx^\mu \wedge {d}x^\nu \qquad {F}%
^{ab}=-F^{ba}.  \label{f207}
\end{equation}
Using (\ref{f203}), (\ref{f204}) and (\ref{f206}) it can be shown that
\begin{equation}
n^aDn^a=0.
\end{equation}

Let the $N$-dimensional Dirac matrix $\gamma _a$ ($a=1,2,\cdots ,{N}$) be
the basis of the Clifford algebra which satisfies
\begin{equation}
\gamma _a\gamma _b+\gamma _b\gamma _a=2\delta _{ab}.  \label{f208}
\end{equation}
An unit vector field $\vec{n}$ on ${\bf M}$ can be expressed as a vector of
Clifford Algebra
\begin{equation}
n=n^a\gamma _a,  \label{f209}
\end{equation}
which can be correspondingly written as
\begin{equation}
n=\frac 1{||\phi ||}\phi ,
\end{equation}
where $\phi $ is a Clifford Algebra vector also
\begin{equation}
\phi =\phi ^a\gamma _{a.}
\end{equation}
The spin connection $1$-form and curvature $2$-form are respectively
represented as Clifford-algebra-valued differential forms
\begin{equation}
\omega =\frac 12\omega ^{ab}I_{ab}\qquad F={\frac 12}F^{ab}I_{ab},
\label{f210}
\end{equation}
in which $I_{ab}$ is the generator of the spin representations of the group $%
SO(N)$
\begin{equation}
I_{ab}={\frac 14}[\gamma _a,\gamma _b]={\frac 14}(\gamma _a\gamma _b-\gamma
_b\gamma _a).  \label{f211}
\end{equation}

By (\ref{f204}) and making use of (\ref{f208}) and (\ref{f209}), it's easy
to prove that the covariant derivative $1$-form of $n^a$ can be represented
in terms of $n$ and $\omega $
\begin{equation}  \label{f212}
Dn=dn-[\omega ,n],
\end{equation}
and curvature $2$-form
\begin{equation}  \label{f213}
F=d\omega -\omega \wedge \omega .
\end{equation}

Arbitrary $U\in Spin(N)$, which satisfies
\begin{equation}
UU^{\dag }=U^{\dag }U=I,  \label{f216}
\end{equation}
is an even-versor\cite{hest}. The induced `spinorial' transformation by $U$
to the basis $\gamma _i$ of the Clifford algebra give $N$ orthonormal
vectors $u_i$\cite{Boer} via
\begin{equation}
u_i:=U\gamma _iU^{\dag }=u_i^a\gamma _a,  \label{f217}
\end{equation}
where $u_i^a$ is the coefficient of $u_i$ in the representation of Clifford
algebra. From the relationship between $U$ and $u_i^a$, we see that $u_i$
has the same singular points with respect to different $``i"$. By (\ref{f217}%
), it is easy to verify that $u_i$ satisfy
\begin{equation}
u_iu_j+u_ju_i=2\delta _{ij},\quad \quad \quad i,j=1,2,\cdots N.  \label{comm}
\end{equation}
From (\ref{f212}) we know that the covariant derivative 1-form of $u_i$ is
\begin{equation}
Du_i=du_i-[\omega ,u_i].  \label{f222}
\end{equation}

There exists the following formula for a Clifford Algebra $r$-vector $A$\cite
{hest}
\begin{equation}
u_iAu_i=(-1)^r(N-2r)A.  \label{comm1}
\end{equation}
For $\omega $ is a Clifford Algebra $2$-vector and using (\ref{comm1}), the
spin connection $\omega $ can be decomposed by $N$ orthonormal vectors $u_i$
as
\begin{equation}
\omega ={\frac 14}(du_iu_i-Du_iu_i).  \label{f225}
\end{equation}
It can be proved that the general decomposition formula (\ref{f225}) has
global property and is independent of the choice of the local coordinates%
\cite{lisheng}.

By choosing the gauge condition
\begin{equation}
Du_i=0
\end{equation}
we can define a generalized pseudo-flat spin connection as
\begin{equation}
\omega _0=\frac 14du_iu_i.  \label{pseudo}
\end{equation}
Suppose there exist $l$ singular points $z_i$ ($i=1,2,\cdots ,l.$) in the
orthonormal vectors $u_j$. One can easily prove that at the normal points of
$u_j$
\begin{equation}
F(\omega _0)=0\quad \quad \quad when\quad x\neq z_i.  \label{fis0}
\end{equation}
For the derivative of $u_i$ at the singular points $z_i$ is undefined, the
formula (\ref{fis0}) is invalid at $z_i$.
\begin{equation}
F\left\{
\begin{array}{c}
=0 \\
\neq 0
\end{array}
\quad \quad \quad when\quad
\begin{array}{c}
x\neq z_i, \\
x=z_i.
\end{array}
\right. .  \label{f235}
\end{equation}
This is why we call $\omega _0$ the pseudo-flat spin connection.

\section{The Gauss-Bonnet-Chern form}

The $GBC$ form is a $N$(even)-form $\Lambda $ over an even dimensional
compact and oriented Riemannian manifold ${\bf M}$, such that when pulled
back to $S({\bf M})$, it's exact
\begin{equation}
\pi ^{*}\Lambda =d\Omega ,  \label{f302}
\end{equation}
where
\begin{equation}
\Lambda =\frac 1{\pi ^{N/2}N!!}Tr(\gamma {F}\wedge {F}\cdots \wedge {F}).
\label{f303}
\end{equation}
Using a recursion method Chern$\cite{Cher1}$ has proved that the $(N-1)$
form $\Omega $ on $S({\bf M})$ is
\begin{equation}
\Omega =\frac 1{(2\pi )^{N/2}}\sum\limits_{k=0}^{N/2-1}(-1)^k\frac{2^{-k}}{%
(N-2k-1)!!k!}\Theta _k,\quad \quad N\geq 4  \label{f304}
\end{equation}
which is called the Chern form with
\begin{equation}
(N-2k-1)!!=(N-2k-1)(N-2k-3)(N-2k-5)\cdots 1,
\end{equation}
\[
\Theta _k=\frac{(-1)^{N/2}}{2^{(N-2k)/2}}Tr(\gamma nDn\wedge \cdots \wedge
Dn\wedge {F\wedge \cdots \wedge F}).
\]
By virtue of the Bianchi identity
\begin{equation}
DF=0,  \label{f306}
\end{equation}
it can be proved that
\begin{equation}
D\Lambda =0.  \label{f307}
\end{equation}
Thus $\Lambda $ determines a cohomology class belonging to the cohomology
groups $H^N({\bf M})$ of ${\bf M}$, which is independent of the connection.
This is equivalent to saying that the integral ${\int_{{\bf M}}}\Lambda $
taken over a closed manifold ${\bf M}$ is a topological invariant\cite{Dubr}%
: the Euler-Poincar\'{e} characteristic $\chi ({\bf M})$. We note that $\pi
^{*}$ maps the cohomology of ${\bf M}$ into that of $S({\bf M})$, while $%
n^{*}$ performs the inverse operation, i.e., $n^{*}\pi ^{*}$ amounts to the
identity. The famous $GBC$ theorem can thus be expressed as
\begin{equation}
\chi ({\bf M})=\int_{{\bf M}}\Lambda =\int_{{\bf M}}n^{*}\pi ^{*}\Lambda
=\int_{{\bf M}}n^{*}d\Omega .  \label{f318}
\end{equation}
Pull back to $S({\bf M})$, (\ref{f318}) becomes
\begin{equation}
\chi ({\bf M})=\int_{n({\bf M})}d\Omega .  \label{f3181}
\end{equation}
When the manifold ${\bf M}$ has boundary, i.e. $\partial {\bf M}\neq 0$,
Gilkey\cite{Gilky} showed that
\begin{equation}
\chi ({\bf M})=\int_{n({\bf M})}d\Omega +\int_{n(\partial {\bf M})}\Omega
\end{equation}

From the well-known Chern-Weil homomorphism\cite{Koba,Nosh}, we know that
the Euler classes is independent of the connection. Hence, we have many
choices of spin connection and the choice depends on the convenience of
calculus. In the present research, we use the generalized pseudo-flat spin
connection to compute the Euler number.

Let $\omega $ takes as the pseudo-flat spin connection (\ref{pseudo}). Then
the curvature 2-form vanishes everywhere but the singular points $z_l$ of $%
u_i.$ It is clearly to see
\begin{equation}
\Omega =\frac{(-1)^{N/2}}{2^N\pi ^{N/2}(N-1)!!}Tr(\gamma nDn\wedge Dn\wedge
\cdots \wedge Dn)
\end{equation}
According to (\ref{comm1}), $n$ can be written as
\begin{equation}
n=n_iu_i,  \label{f320}
\end{equation}
where $n_i$ is the projection of $n$ on $u_i$
\begin{equation}
n_i=\frac 12(nu_i+u_in).  \label{f321}
\end{equation}
Noticing that $u_i$ and $du_i$ are both Clifford Algebra vectors, using (\ref
{comm1}) we have
\begin{equation}
u_ju_iu_j=-(N-2)u_i,\quad \quad u_jdu_iu_j=-(N-2)du_i.  \label{d3a}
\end{equation}
Then $Dn$ becomes
\begin{equation}
Dn=dn_iu_i.  \label{f322}
\end{equation}
As a result
\begin{equation}
\Omega =\frac{(-1)^{N/2}}{2^N\pi ^{N/2}(N-1)!!}Tr(\gamma
u_{i_1}u_{i_2}\cdots u_{i_N})n_{i_1}dn_{i_2}\wedge \cdots \wedge dn_{i_N}.
\label{f323a}
\end{equation}
It can be deduced that $\Omega $ is
\begin{equation}
\Omega =\frac 1{(n-1)!A(S^{N-1})}\epsilon _{a_1a_2\cdots
a_N}n_{a_1}dn_{a_2}\wedge \cdots \wedge dn_{a_N}{.}  \label{f323}
\end{equation}
where $A(S^{N-1})$ is the area of $S^{N-1}$
\begin{equation}
A(S^{N-1})={\frac{2\pi ^{N/2}}{{\Gamma (\frac N2)}}}.
\end{equation}

For $n_a$ are just the projections of $n$ on the basis $u_a$, we have
\begin{equation}
n_an_a=1.  \label{f324}
\end{equation}
The above expression (\ref{f323}) is nothing but the $GBC$-form expressed
cleanly in terms of the unit vector $n$.

\section{Topological structure of $GBC$ density on the manifold without
boundary}

We can write $n_a$ as
\begin{equation}
n_a=\frac 12(nu_a+u_an)=\frac{\frac 12(\phi u_a+u_a\phi )}{||\phi ||}=\frac{%
\phi _a}{||\phi ||},  \label{f325}
\end{equation}
where $\phi _a$ are the projections of $\phi $ on $u_{\left( a\right) }$, so
it is easy to prove that
\begin{equation}
||\phi ||=\sqrt{\phi ^a\phi ^a}=\sqrt{\phi _a\phi _a}.  \label{f327}
\end{equation}
The derivative of $n_a$ can be deduced as
\begin{equation}
dn_a=\frac{d\phi _a}{||\phi ||}-\phi _ad({\frac 1{||\phi ||}}).
\end{equation}
Substituting it into (\ref{f323}), we have the expression of $\Omega $ on $S(%
{\bf M})$
\begin{equation}
\Omega =\frac 1{A(S^{N-1})(N-1)!}\epsilon _{a_1a_2\cdots a_N}\frac{\phi
_{a_1}}{||\phi ||^N}d\phi _{a_2}\wedge \cdots \wedge d\phi _{a_N}.
\label{f3271}
\end{equation}
Using
\begin{equation}
\frac{\phi _a}{||\phi ||^N}=-\frac 1{N-2}\frac \partial {\partial \phi _a}(%
\frac 1{||\phi ||^{N-2}}),
\end{equation}
the pull back of the exterior derivative of (\ref{f3271}) to ${\bf M}$ can
be written as
\begin{eqnarray}
n^{*}d\Omega  &=&-\frac 1{A(S^{N-1})(N-1)!(N-2)}\epsilon _{a_1a_2\cdots a_N}%
\frac \partial {\partial \phi _a}\frac \partial {\partial \phi _{a_1}}(\frac %
1{||\phi ||^{N-2}})  \nonumber \\
&&\times \frac{\partial \phi _a}{\partial {x}^{\mu _1}}\frac{\partial \phi
_{a_1}}{\partial {x}^{\mu _2}}\cdots \frac{\partial \phi _{a_N}}{\partial {x}%
^{\mu _N}}\frac{\epsilon ^{\mu _1\mu _2\cdots \mu _N}}{\sqrt{g}}\sqrt{g}d^Nx,
\end{eqnarray}
where $g=det(g_{\mu \nu })$, $g_{\mu \nu }$ is the metric tensor of ${\bf M}$%
. Define the Jacobian $D(\phi /x)$ as
\begin{equation}
\epsilon _{a_1a_2\cdots {a_N}}D(\phi /x)=\epsilon ^{\mu _1\mu _2\cdots \mu
_N}\partial _{\mu _1}\phi _{a_1}\partial _{\mu _2}\phi _{a_2}\cdots \partial
_{\mu _N}\phi _{a_N}.
\end{equation}
Noticing
\begin{equation}
\epsilon _{a_1a_2\cdots a_N}\epsilon _{aa_2\cdots a_N}=(N-1)!\delta _{a_1a},
\end{equation}
we get
\begin{equation}
n^{*}d\Omega =-\frac 1{A(S^{N-1})(N-2)}\frac{\partial ^2}{\partial \phi
_a\partial \phi _a}(\frac 1{||\phi ||^{N-2}})D(\frac \phi x)d^Nx.
\end{equation}
The general Green's-function formula$\cite{Gelf}$ in $\phi $ space is
\begin{equation}
\triangle _\phi (\frac 1{||\phi ||^{N-2}})=-\frac{4\pi ^{N/2}}{\Gamma (N/2-1)%
}\delta (\phi )\quad \quad N\geq 3,  \label{f553}
\end{equation}
where
\begin{equation}
\triangle _\phi =\frac{\partial ^2}{\partial \phi _a\partial \phi _a},
\end{equation}
is the $N$-dimensional Laplacian operator in $\phi $ space. We obtain the
new formulation of $GBC$ form in terms of $\delta $ function $\delta (\phi )$
\begin{equation}
n^{*}d\Omega =\delta (\phi )D(\phi /x)d^Nx.  \label{f328}
\end{equation}

Since $\phi ^a(x)$ has $l$ isolated zeroes on ${\bf M}$ and let the $i$th
zero be $z_i$, it is well known from the ordinary theory of the $\delta $%
-function \cite{Schw} that
\begin{equation}
\delta (\phi )=\sum\limits_{i=1}^l\frac{\beta _i\delta (x-z_i)}{D(\phi
/x)|_{x=z_i}}.  \label{f330}
\end{equation}
Then one obtains
\begin{equation}
\delta (\phi )D(\frac \phi x)=\sum\limits_{i=1}^l\beta _i\eta _i\delta
(x-z_i),  \label{f331}
\end{equation}
where $\beta _i$ is the positive integer (the Hopf index of the $i$th zero)
and $\eta _i$ the Brouwer degree\cite{Dubr,Miln1}
\begin{equation}
\eta _i=sgnD(\phi /x)|_{x=z_i}=\pm 1.
\end{equation}
From the above deduction the following topological structure is obtained:
\begin{equation}
n^{*}d\Omega =\delta (\phi )D(\frac \phi x)d^Nx=\sum\limits_{i=1}^l\beta
_i\eta _i\delta (x-z_i)d^Nx,  \label{f332}
\end{equation}
which means that the local structure of $n^{*}d\Omega _0$ is labeled by the
Brouwer degrees and Hopf indices, which are topological invariants.
Therefore the Euler-Poincar\'{e} characteristic $\chi (M)$ can be
represented as
\begin{equation}
\chi ({\bf M})=\int_{{\bf M}}n^{*}d\Omega =\sum\limits_{i=1}^l\beta _i\eta
_i.  \label{f333}
\end{equation}

On another hand the above formula also gives the winding number of the
manifold $M$ and the mapping $\vec{\phi}$ (see, for e.g [])
\[
W(\phi ,z_i)=\beta _i\eta _i.
\]
{\em \ }Then the Euler-Poincar\'{e} characteristic $\chi ({\bf M})$ can
further be expressed in terms of winding numbers and degree of $\phi $%
\begin{equation}
\chi ({\bf M})=\deg \phi =W(\phi ,{\bf M})=\sum_{i=1}^lW(\phi ,z_i).
\end{equation}

From (\ref{f201}), we know that the zeroes of $\phi $ are just the
singularities of $n$. Here (\ref{f333}) says that the sum of the indices of
the singular points of $n,$ or of the zeroes of $\phi $, is the
Euler-Poincar\'e characteristic. Therefore the topological structure of $GBC$
density reveals the expected result of the Hopf theorem. The above
discussions, especially, the expressions (\ref{f332}) is very valuable to
establish the theory of the $GBC$ topological current. Since the $GBC$
theorem is also correct for a pseudo-Riemannian manifold\cite{Cher3}, from (%
\ref{f332}) we know that the $GBC$ density is related to instantons for
Einstein space-time.

\section{Topological structure of $GBC$ density on the manifold with boundary
}

When we take a manifold ${\bf M}$ with boundary $\partial {\bf M}$, the
Euler characteristic becomes
\begin{equation}
\chi =\int_{n({\bf M})}d\Omega +\int_{n(\partial {\bf M)}}\Omega
\label{euler-boun}
\end{equation}
The discussion of the first integral term on the left-hand side of the above
equation keeps the same as what we discussed in the manifold without
boundary
\begin{equation}
\int_{n({\bf M})}d\Omega =\sum_{i=1}^l\int_{{\bf M}}\beta _i\eta _i\delta
(x-z_i)d^nx=\sum_{i=1}^l\beta _i\eta _i
\end{equation}
Hence, what we need to discuss is the second integral term on the right-hand
side of equation (\ref{euler-boun}).

With the same reason as mentioned before, this integral terms is induced as
\begin{equation}
\int_{n(\partial {\bf M)}}\Omega =\int_{\partial {\bf M}}\frac 1{%
(n-1)!A(S^{N-1})}\epsilon _{a_1a_2\cdots a_N}n_{a_1}dn_{a_2}\wedge \cdots
\wedge dn_{a_N}{.}
\end{equation}
Denote $m_{\bot }^a$ as the unit vector orthonormal to the boundary of ${\bf %
M}$, i.e.
\begin{equation}
m_{\bot }^a\bot \partial {\bf M}
\end{equation}
and $m_{//}^a$ be the unit vector which is parallel to $\partial {\bf M}$.
Now $n$ can be represented by $m_{\bot }^a$ and $m_{//}^a$%
\begin{equation}
n^a=m_{\bot }^a\cos \alpha +m_{//}^a\sin \alpha .
\end{equation}
The projection of $\phi $ on the boundary $\partial {\bf M}$ is denoted as $%
\phi _{//}$%
\begin{equation}
\phi _{//}^a=\phi ^a-m_{\bot }^a\phi ^bm_{\bot }^b
\end{equation}
By choosing the direction of the basis of the vector field, we can make $%
m_{\bot }$ taking as $(0,\cdots 0,1)$ then it can proved $m_{//}$ takes the
form
\begin{equation}
m_{//}=(m^1,\cdots ,m^{N-1},0)
\end{equation}
and
\begin{equation}
m^Am^A=1\quad \quad \quad A=1,2,\cdots ,N-1
\end{equation}
Then we have
\begin{equation}
n=(m^A\sin \alpha ,\cos \alpha )
\end{equation}
and
\begin{equation}
\phi _m=(\phi _{//}^1,\cdots ,\phi _{//}^{N-1},0)
\end{equation}
Using this formula we can in forth to express the Chern-density as
\begin{eqnarray}
\Omega &=&\frac 1{(n-1)!A(S^{N-1})}\epsilon _{A_1A_2\cdots {A_N}}(\sin
^{N-1}\alpha \cos \alpha dm^{A_1}\wedge \cdots \wedge {dm}^{A_{N-1}}
\nonumber \\
&&+(N-1)m^{A_1}\sin ^{N-2}\alpha d\alpha \wedge dm^{A_2}\wedge \cdots \wedge
{dm}^{A_{N-1}})
\end{eqnarray}
which can be written as the sum of two parts
\begin{equation}
\Omega =\rho _1+\rho _2
\end{equation}
and
\begin{eqnarray}
\rho _1 &=&\frac{\epsilon _{A_1A_2\cdots A_N}}{(N-1)!A(S^{N-1})}\sin
^{N-1}\alpha \cos \alpha dm^{A_1}\wedge \cdots \wedge {dm}^{A_{N-1}} \\
\rho _2 &=&\frac{\epsilon _{A_1A_2\cdots A_N}}{(N-1)!A(S^{N-1})}m^{A_1}\sin
^{N-2}\alpha d\alpha \wedge dm^{A_2}\wedge \cdots \wedge {dm}^{A_{N-1}}
\end{eqnarray}
By choice a term of coordinate $(u^1,\cdots ,u^{N-1},v)$ on the boundary $%
\partial {\bf M}$ and let $u=(u^1,\cdots ,u^{N-1})$ be the inner coordinate
of $\partial {\bf M}$. Using the same method in discussing the Chern
density, we can get $\rho _1$ to be
\begin{equation}
\rho _1=\frac 1{(n-1)!A(S^{N-1})}\sin ^{N-1}\alpha \cos \alpha \delta
^{N-1}(\phi _{//})d^{N-1}u
\end{equation}

Notice that only at the zeroes of $\phi _{//}$ the delta function does not
vanish. However at the zeroes of $\phi _{//}$ the angle $\alpha $ is $0$ or $%
\pi $. Then
\begin{equation}
\sin ^{N-1}\alpha \cos \alpha =0
\end{equation}
Hence, $\rho _1$ contribute nothing to the Euler number for at the zeroes of
$\phi _{//}$.

Now we turn to study the properties of $\rho _2$. For the continuous of $%
\phi _{//}$, there must exist a closed $N-2$-dimensional hypersurface $%
S^{N-2}(\alpha )$ in $\partial {\bf M}$, on which the angle $\alpha $ keeps
the same. Therefore we can divide the integral of $\rho _2$ as
\begin{eqnarray}
\int_{\partial {\bf M}}\rho _2 &=&\frac{\epsilon _{{A}_1\cdots A_{N-1}}}{%
2(N-2)!A(S^{N-1})}\int_0^\pi \sin ^{N-2}\alpha d\alpha \int_{s(\alpha
)}m^{A_1}dm^{A_2}\wedge \cdots \wedge {dm}^{A_{N-1}}  \nonumber \\
&=&\frac{\epsilon _{{A}_1\cdots A_{N-1}}}{2(N-2)!A(S^{N-2})}\int_{s(\alpha
)}m^{A_1}dm^{A_2}\wedge \cdots \wedge {dm}^{A_{N-1}}
\end{eqnarray}
in which we use the integral formula
\begin{equation}
\int_0^\pi \sin ^{N-2}\alpha d\alpha =\sqrt{\pi }\frac{\Gamma (\frac{N-1}2)}{%
\Gamma (\frac N2)}
\end{equation}
and the factor $\frac 12$ comes from that the integral limits $\alpha =0$
and $\alpha =\pi $ should be seen as one points, which is computed twice.
Therefore we must insert the factor $\frac 12$ to make the compute correct.
Denote the $N-1$\thinspace dimensional super surface enclosed by $%
S^{N-2}(\alpha )$ as $S^{N-1}(\alpha )$ then from the stoke's theorem we
have
\begin{equation}
\int_{\partial {\bf M}}\rho _2=\frac 1{2(N-2)!A(S^{N-2})}\epsilon _{{A}%
_1\cdots A_{N-1}}d\alpha \int_{s(\alpha )}dm^{A_1}\wedge dm^{A_2}\wedge
\cdots \wedge {dm}^{A_{N-1}}
\end{equation}
Now it easy to proved analogous to that was proved before that
\begin{eqnarray}
\int_{\partial {\bf M}}\rho _2 &=&\frac 12\int_{\partial {\bf M}}\delta
(\phi _{//})J(\frac{\phi _{//}}u)d^{N-1}u  \nonumber \\
&=&\frac 12\sum_{l=1}^k\beta _l(\phi _{//})\eta _l(\phi _{//})=\frac 12\deg
\phi _{//}
\end{eqnarray}
where $k$ is the number of the zeroes of $\phi _{//}$ and $\beta _l(\phi
_{//}),$ $\eta _l(\phi _{//})$ are the Brouwer degree and Hopf index of $%
\phi _{//}$ on the $l$-th zero of $\phi _{//}$. Finally we obtain the index
theorem of the manifold ${\bf M}$ with boundary $\partial {\bf M}$%
\begin{equation}
\chi ({\bf M})=\sum_{i=1}^l\beta _i(\phi )\eta _i(\phi )+\frac 12%
\sum_{i=1}^k\beta _i(\phi _{//})\eta _i(\phi _{//}).
\end{equation}
In terms of winding umbers and indices, it is
\begin{eqnarray}
\chi ({\bf M}) &=&\sum_{i=1}^lW_i(\phi )+\frac 12\sum_{i=1}^kW_i(\phi _{//})
\nonumber \\
&=&\deg \phi +\frac 12\deg \phi _{//}
\end{eqnarray}
This formula give a revise of the Hopf theorem on the manifold with
boundary. If the angle $\alpha $ keeps invariant, especially when $\alpha =0$%
, i.e. $\phi $ is parallel to $\partial {\bf M}$, One can find that the
boundary term vanishes, and
\begin{equation}
\chi ({\bf M})=\sum_{i=1}^lW_i(\phi )=\deg \phi .
\end{equation}

\section{Conclusion}

In conclusion, we have explicitly constructed the general decomposition
theory of the spin connection of the group $SO(N)$ by virtue of $N$
orthonormal vectors on the compact and oriented Riemannian manifold via
Clifford algebra. We have proved the Euler-Poincar\'{e} characteristic $\chi(%
{\bf M})$ of a manifold ${\bf M} $ with boundary $\partial {\bf M} $ is
equal to the sum of the total index of a smooth vector field $\phi$ on ${\bf %
M}$ and half the total index of the projective vector field of $\phi$ on $%
\partial {\bf M}$. The boundary term vanishes when the vector field is
always transversal to the boundary and pointed outwards. Our results
indicate that the Hopf indices and Brouwer degrees label the local structure
of the Euler density. The Euler-Poincar\'{e} characteristic relates to the
index, or equivalently the winding number, of the vector filed $\phi$.

\end{document}